\documentclass{cpbtex}

\begin{document}
\begin{CJK*}{GBK}{song}

\title{Construction of multiple soliton solutions of the quintic nonlinear Schr\"odinger equation\thanks{Project supported by the National Natural Science Foundation of China (Grant No.~11975145).}}


\author{Zhou-Zheng Kang$^{1,2}$, \ Tie-Cheng Xia$^{1}\thanks{Corresponding author: xiatc@shu.edu.cn}$\\
\small{$^{1}${Department of Mathematics, Shanghai University, Shanghai 200444, China}}\\  
\small{$^{2}${College of Mathematics and Physics, Inner Mongolia University for Nationalities, Tongliao 028043, China}} 
}   


\date{\today}
\maketitle

\begin{abstract}
In this paper, an extended nonlinear Schr\"odinger equation with higher-order that includes fifth-order dispersion with matching higher-order nonlinear
terms is investigated under zero boundary condition at infinity. Carrying out the spectral analysis, a kind of matrix Riemann-Hilbert problem is formulated on the real axis. Then on basis of the resulting matrix Riemann-Hilbert problem under restriction of no reflection, multiple soliton solutions of the extended nonlinear Schr\"odinger equation are generated explicitly.
\end{abstract}



\section{Introduction}
Soliton solutions of nonlinear evolution equaitons (NLEEs) play an especially important significance in studying a variety of complex nonlinear phenomena in fluid dynamics, plasma physics, oceanography, optics, condensed matter physics and so forth.
By now, many efficient approaches have been available for finding soliton solutions, some of which include inverse scattering transformation [1,2], Darboux transformation [3--6], B\"acklund transformation [7], Riemann-Hilbert method [8], and Hirota's bilinear method [9--12]. In recent years, there has been an increasing interest in exploring abundant multiple soliton solutions of NLEEs via the Riemann-Hilbert method, including the coupled derivative Schr\"odinger equation [13], the Kundu-Eckhaus equation [14], and others
[15--21].

In this work, under zero boundary condition at infinity, we would like to consider the quintic nonlinear Schr\"odinger equation
\begin{equation}
i{{q}_{t}}+\frac{1}{2}{{q}_{xx}}+{{\left| q \right|}^{2}}q-i\varepsilon \big({{q}_{xxxxx}}+10{{\left| q \right|}^{2}}{{q}_{xxx}}+30{{\left| q \right|}^{4}}{{q}_{x}}+20{{q}^{*}}{{q}_{x}}{{q}_{xx}}+10{{\big(q{{\left| {{q}_{x}} \right|}^{2}}\big)}_{x}}\big)=0,
\end{equation}
where $\left| q \right|$ denotes the envelope of the waves, and $x$ is the propagation variable, and $t$
is the transverse variable (time in a moving frame). $\varepsilon$ is a real parameter.
Several studies have been conducted.
Chowdury et al. [22] showed that a breather solution of Eq. (1) can be changed into a nonpulsating soliton solution on a background. And locus of the eigenvalues on the complex plane which convert breathers into solitons was worked out. They also studied the interaction between the resulting solitons, as well as between breathers and these solitons.
The superregular
breather, multi-peak soliton and hybrid solutions were investigated by Wang et al. [23] via the modified Darboux transformation and Joukowsky transform.

\section{Desired Lax pair}
The Lax pair [23] for Eq. (1) reads as
\begin{subequations}
\begin{align}
 {{\psi }_{x}}& =U\psi =(\lambda {{U}_{0}}+Q)\psi , \\
 {{\psi }_{t}}& =V\psi =(V_{1}+V_{2})\psi ,
\end{align}
\end{subequations}
where $\psi =({{\psi }_{1}},{{\psi }_{2}})^{T}$
is the spectral function,
$${{U}_{0}}=\left( \begin{matrix}
   -i & 0  \\
   0 & i  \\
\end{matrix} \right),\quad Q=\left( \begin{matrix}
   0 & q  \\
   -{{q}^{*}} & 0  \\
\end{matrix} \right),\quad V_{1}=\left( \begin{matrix}
   A & 0  \\
   0 & -A  \\
\end{matrix} \right),\quad V_{2}=\left( \begin{matrix}
   0 & B  \\
   -{{B}^{*}} & 0  \\
\end{matrix} \right),$$
$\lambda\in\mathbb{C}$ is a spectral parameter, and
\begin{align*}
&A=-16i\varepsilon {{\lambda }^{5}}+8i\varepsilon {{\lambda }^{3}}{{\left| q \right|}^{2}}+4\varepsilon {{\lambda }^{2}}(qq_{x}^{*}-{{q}_{x}}{{q}^{*}})-i{{\lambda }^{2}}-2i\varepsilon \lambda \big(qq_{xx}^{*}+{{q}^{*}}{{q}_{xx}}-{{\left| {{q}_{x}} \right|}^{2}}+3{{\left| q \right|}^{4}}\big) \\
&\quad\quad+\frac{1}{2}i{{\left| q \right|}^{2}}+\varepsilon \big({{q}^{*}}{{q}_{xxx}}-qq_{xxx}^{*}+{{q}_{x}}q_{xx}^{*}-{{q}_{xx}}q_{x}^{*}+6{{\left| q \right|}^{2}}{{q}^{*}}{{q}_{x}}-6{{\left| q \right|}^{2}}q_{x}^{*}q\big),\\
&B=16\varepsilon {{\lambda }^{4}}q+8i\varepsilon {{\lambda }^{3}}{{q}_{x}}-4\varepsilon {{\lambda }^{2}}\big({{q}_{xx}}+2{{\left| q \right|}^{2}}q\big)-2i\varepsilon \lambda \big({{q}_{xxx}}+6{{\left| q \right|}^{2}}{{q}_{x}}\big)+\lambda q \\
&\quad\quad+\varepsilon \big({{q}_{xxxx}}+8{{\left| q \right|}^{2}}{{q}_{xx}}+2{{q}^{2}}q_{xx}^{*}+4{{\left| {{q}_{x}} \right|}^{2}}q+6q_{x}^{2}{{q}^{*}}+6{{\left| q \right|}^{4}}q\big)+\frac{1}{2}i{{q}_{x}}.
\end{align*}

We first need to convert the above Lax pair into the equivalent form
\begin{align}
 {{\psi }_{x}}& =(-i\lambda \Lambda +Q)\psi , \\
 {{\psi }_{t}}& =((-16i\varepsilon {{\lambda }^{5}}-i{{\lambda }^{2}})\Lambda +\tilde{V})\psi ,
\end{align}
in which $\Lambda =\text{diag}(1,-1),$
\begin{align*}
&\tilde{V}=\Big(8i\varepsilon {{\lambda }^{3}}{{\left| q \right|}^{2}}+4\varepsilon {{\lambda }^{2}}(qq_{x}^{*}-{{q}_{x}}{{q}^{*}})-2i\varepsilon \lambda \big(qq_{xx}^{*}+{{q}^{*}}{{q}_{xx}}-{{\left| {{q}_{x}} \right|}^{2}}+3{{\left| q \right|}^{4}}\big)+\frac{1}{2}i{{\left| q \right|}^{2}} \\
&\quad\quad+\varepsilon \big({{q}^{*}}{{q}_{xxx}}-qq_{xxx}^{*}+{{q}_{x}}q_{xx}^{*}-{{q}_{xx}}q_{x}^{*}+6{{\left| q \right|}^{2}}{{q}^{*}}{{q}_{x}}-6{{\left| q \right|}^{2}}q_{x}^{*}q\big)\Big)\Lambda+V_{2},
\end{align*}

Under the hypothesis of $q\rightarrow0$ as $x\rightarrow\pm\infty$, we can see from (3)--(4) that $\psi \propto {{e}^{-i\lambda \Lambda x-(16i\varepsilon {{\lambda }^{5}}+i{{\lambda }^{2}})\Lambda t}}$. Hence, a variable transformation
\begin{equation*}
\psi =J{{e}^{-i\lambda \Lambda x-(16i\varepsilon {{\lambda }^{5}}+i{{\lambda }^{2}})\Lambda t}},
\end{equation*}
is employed to change (2) and (3) into the desired form
\begin{align}
 {{J}_{x}}& =-i\lambda [\Lambda ,J]+QJ, \\
 {{J}_{t}}& =(-16i\varepsilon {{\lambda }^{5}}-i{{\lambda }^{2}})[\Lambda ,J]+\tilde{V}J.
\end{align}
\section{Matrix Riemann--Hilbert problem}
The matrix Jost solutions can be expressed into a collection of columns
\begin{equation}
{{J}_{-}}=([{{J}_{-}}]_{1},[{{J}_{-}}]_{2}),\quad{{J}_{+}}=([{{J}_{+}}]_{1},[{{J}_{+}}]_{2}),
\end{equation}
having the asymptotic conditions
\begin{equation}
\begin{aligned}
 & {{J}_{-}}\to \mathbf{I_{2}},\quad x\to -\infty , \\
 & {{J}_{+}}\to \mathbf{I_{2}},\quad x\to +\infty ,
\end{aligned}
\end{equation}
in which $\mathbf{I_{2}}$ stands for the identity matrix of rank 2.
As a matter of fact, $J_{\pm}(x,\lambda )$ are uniquely determined by the solution of Volterra integral equations
\begin{subequations}
\begin{align}
 & {{J}_{-}}(x,\lambda )=\mathbf{I_{2}}+\int_{-\infty }^{x}{{{e}^{-i\lambda \Lambda (x-y )}}Q(y){{J}_{-}}(y,\lambda ){{e}^{i\lambda \Lambda (x-y )}}dy }, \\
 & {{J}_{+}}(x,\lambda )=\mathbf{I_{2}}-\int_{x}^{+\infty }{{{e}^{-i\lambda \Lambda (x-y )}}Q(y){{J}_{+}}(y,\lambda ){{e}^{i\lambda \Lambda (x-y )}}dy },
\end{align}
\end{subequations}

Then, Eqs. (9a) and (9b) are analyzed to indicate that $[{{J}_{-}}]_{1},[{{J}_{+}}]_{2}$ allow analytical extensions to $\mathbb{C_{+}}$, however $[{{J}_{+}}]_{1},[{{J}_{-}}]_{2}$ are analytically extendible to $\mathbb{C_{-}}$, where
\begin{equation*}
\mathbb{C_{-}}=\{\lambda\in\mathbb{C}|\textrm{Im}(\lambda)<0\},\quad\mathbb{C_{+}}=\{\lambda\in\mathbb{C}|\textrm{Im}(\lambda)>0\},
\end{equation*}

Because ${{J}_{-}}{{e}^{-i\lambda \Lambda x}}$ and ${{J}_{+}}{{e}^{-i\lambda \Lambda x}}$ are both the fundamental matrix solutions of (3), they must be linearly dependent by the scattering matrix $S(\lambda )=(s_{jl})_{2\times2}$, i.e.,
\begin{equation}
{{J}_{-}}{{e}^{-i\lambda \Lambda x}}={{J}_{+}}{{e}^{-i\lambda \Lambda x}}S(\lambda ),\quad \lambda\in\mathbb{R}.
\end{equation}

In consideration of the analytic property of $J_{\pm}$, we define the analytic function in $\mathbb{C_{+}}$ as
\begin{equation}
{{P}_{1}}=({{[{{J}_{-}}]_{1}}},{{[{{J}_{+}}]_{2}}})={{J}_{-}}{{H}_{1}}+{{J}_{+}}{{H}_{2}},
\end{equation}
where
\begin{equation}
{{H}_{1}}=\text{diag}(1,0),\quad{{H}_{2}}=\text{diag}(0,1).
\end{equation}

In what follows, we examine the large-$\lambda$ asymptotic behavior of ${{P}_{1}}$. We expand ${{P}_{1}}$ as
\[{{P}_{1}}=P_{1}^{(0)}+{\lambda}^{-1}{{P}_{1}^{(1)}}+{\lambda}^{-2}{{P}_{1}^{(2)}}+O\big({\lambda}^{-3}\big),\quad\lambda \to \infty ,\]
and carry this expansion into (3). Comparing the coefficients of the same power of $\lambda$ gives rise to
\[\begin{aligned}
& O(1):P_{1x}^{(0)}=-i[\Lambda ,P_{1}^{(1)}]+QP_{1}^{(0)}, \\
& O(\lambda ):0=-i[\Lambda ,P_{1}^{(0)}].
\end{aligned}\]
We can see that
\[{{P}_{1}}\to \mathbf{I_{2}},\quad\lambda \to \infty .\]

For formulating the desired matrix Riemann--Hilbert problem, it is required to construct the other analytic function $P_{2}$. Now we consider the adjoint equation associated with (5)
\begin{equation}
{{\chi }_{x}}=-i\lambda [\Lambda ,\chi ]-\chi Q.
\end{equation}
The inverse matrix of $J_{\pm}^{-1}$ are expressible into
\begin{equation}
{{J}_{-}^{-1}}=\left( \begin{matrix}
   {{[{{J}_{-}^{-1}}]^{1}}}  \\
   {{[{{J}_{-}^{-1}}]^{2}}}  \\
\end{matrix} \right),\quad{{J}_{+}^{-1}}=\left( \begin{matrix}
   {{[{{J}_{+}^{-1}}]^{1}}}  \\
   {{[{{J}_{+}^{-1}}]^{2}}}  \\
\end{matrix} \right),
\end{equation}
where $[{{J}_{\pm}^{-1}}]^{j}(j=1,2)$ denote the $j$-th row of ${J}_{\pm}^{-1}$. It can be verified that $J_{\pm}^{-1}$ meet Eq. (13). Then from Eq. (10), we immediately obtain
\begin{equation}
{{J}_{-}^{-1}}={{e}^{-i\lambda \Lambda x}}S^{-1}(\lambda ){{e}^{i\lambda \Lambda x}}{{J}_{+}^{-1}},
\end{equation}
where ${{S}^{-1}}(\lambda )=(r_{jl})_{2\times2}$.
Thus, the matrix function $P_{2}$ can be expressed in the form
\begin{equation}
{{P}_{2}}=\left( \begin{matrix}
   {{[J_{-}^{-1}]^{1}}}  \\
   {{[J_{+}^{-1}]^{2}}}  \\
\end{matrix} \right)={{H}_{1}}J_{-}^{-1}+{{H}_{2}}J_{+}^{-1},
\end{equation}
with ${{H}_{1}}$ and ${{H}_{2}}$ being given by (12). Moreover, the asymptotic behavior for $P_{2}$ is
\[{{P}_{2}}\to \mathbf{I_{2}},\quad\lambda \to \infty .\]

Inserting (4) into (7) yields
$${{[{{J}_{+}}]_{2}}}={{r}_{12}}{{e}^{-2i\lambda x}}{{[{{J}_{-}}]_{1}}}+{{r}_{22}}{{[{{J}_{-}}]_{2}}}.$$
Therefore, ${{P}_{1}}$ is rewritten as
\begin{equation*}
{{P}_{1}}=({{[{{J}_{-}}]_{1}}},{{[{{J}_{+}}]_{2}}})=({{[{{J}_{-}}]_{1}}},{{[{{J}_{-}}]_{2}}})\left( \begin{matrix}
   1 & {{r}_{12}}{{e}^{-2i\lambda x}}  \\
   0 & {{r}_{22}}  \\
\end{matrix} \right).
\end{equation*}

Carrying (11) into (12) leads to
$${{[J_{+}^{-1}]^{2}}}={{s}_{21}}{{e}^{2i\lambda x}}{{[J_{-}^{-1}]^{1}}}+{{s}_{22}}{{[J_{+}^{-1}]^{2}}}.$$
Then, ${{P}_{2}}$ takes the form
\begin{equation*}
{{P}_{2}}=\left( \begin{matrix}
   {{[J_{-}^{-1}]^{1}}}  \\
   {{[J_{+}^{-1}]^{2}}}  \\
\end{matrix} \right)=\left( \begin{matrix}
   1 & 0  \\
   {{s}_{21}}{{e}^{2i\lambda x}} & {{s}_{22}}  \\
\end{matrix} \right)\left( \begin{matrix}
   {{[J_{-}^{-1}]^{1}}}  \\
   {{[J_{-}^{-1}]^{2}}}  \\
\end{matrix} \right),
\end{equation*}

Based on the above results, a kind of Riemann--Hilbert problem for Eq. (1) on the real axis can be stated as follows
\begin{equation}
{{P}^{-}}(x,\lambda){{P}^{+}}(x,\lambda)=G(x,\lambda),\quad \lambda\in\mathbb{R},
\end{equation}
in which
$$G(x,\lambda)=\left( \begin{matrix}
   1 & {{r}_{12}}{{e}^{-2i\lambda x}}  \\
   {{s}_{21}}{{e}^{2i\lambda x}} & 1  \\
\end{matrix} \right),$$
and ${{s}_{21}}{{r}_{12}}+{{s}_{22}}{{r}_{22}}=1$. The normalization conditions are given by
\begin{align*}
  &{P_{1}}(x,\lambda)\to {\mathbf{I}}_{2},\quad \lambda \in {\mathbb{C}_{+}}\to \infty , \\
  &{P_{2}}(x,\lambda)\to {\mathbf{I}}_{2},\quad \lambda \in {\mathbb{C}_{-}}\to \infty .
\end{align*}

\section{Soliton solutions}
Our focus in this section will be on generating soliton solutions to Eq. (1) on basis of the obtained Riemann--Hilbert problem.
Suppose that the Riemann--Hilbert problem (17) is irregular, which reveals that both $\det {P_{1}}$ and $\det {P_{2}}$ have some zeros in their own analytic domains.
According to the definitions of ${P_{1}}$ and ${P_{2}}$ as well as the scattering relation (7), we have
$$\det {{P}_{1}}(\lambda)={{r}_{22}}(\lambda),\quad\det {{P}_{2}}(\lambda)={{s}_{22}}(\lambda),$$
which show that $\det {{P}_{1}}$ and $\det {{P}_{2}}$ are in possession of the same zeros as ${{r}_{22}}$ and ${{s}_{22}}$ respectively.

In what follows, we need to specify the zeros. Manifestly, the matrix $Q$ is skew-Hermitian,
$
Q^{\dagger }=-Q.
$
On basis of this property, we deduce that
\begin{equation}
J_{\pm }^{\dagger }({{\lambda}^{*}})=J_{\pm }^{-1}(\lambda).
\end{equation}
Taking the Hermitian of Eq. (11) and using Eq. (18), we have
\begin{equation}
P_{1}^{\dagger }({{\lambda}^{*}})={{P}_{2}}(\lambda),
\end{equation}
and
\begin{equation}{{S}^{\dagger }}({{\lambda}^{*}})={{S}^{-1}}(\lambda),\end{equation}
for $\lambda \in {\mathbb{C}^{-}}.$ From Eq. (20), we further find
\begin{equation}
s_{22}^{*}({{\lambda}^{*}})={{r}_{22}}(\lambda),\quad \lambda \in {\mathbb{C}^{+}}.
\end{equation}
Therefore, we assume that $\det {{P}_{1}}$ has $N$ simple zeros $\lambda_{j}$ in $\mathbb{C}^{+}$ and $\det {{P}_{2}}$ has $N$ simple zeros $\hat{\lambda}_{j}$ in $\mathbb{C}^{-}$, where $\hat{\lambda}_{j}=\lambda^{\ast}_{j}$. Each of $\ker {P_{1}}({{\hat{\lambda }}_{j}})$ includes only a single basis column vector ${{\omega}_{j}}$, and each of $\ker {P_{2}}({{\hat{\lambda }}_{j}})$ includes only a single basis row vector ${{\hat{\omega}}_{j}}$,
\begin{align}
&{P_{1}}({{\lambda}_{j}}){{\omega}_{j}}=0,\\
&{{\hat{\omega}}_{j}}{P_{2}}({{\hat{\lambda}}_{j}})=0.
\end{align}
Taking the Hermitian of Eq. (22) and using (19), we find that the eigenvectors
fulfill the relation
\begin{equation}
{{\hat{\omega}}_{j}}=\omega_{j}^{\dagger },\quad 1\le j\le N.
\end{equation}
Taking the $x$-derivative and $t$-derivative of Eq. (22) respectively and using (3)--(4), we arrive at

$${{P}_{1}}({{\lambda}_{j}})\left( \frac{\partial {{\omega}_{j}}}{\partial x}+i{{\lambda}_{j}}\Lambda {{\omega}_{j}} \right)=0,\quad
{{P}_{1}}({{\lambda}_{j}})\left( \frac{\partial {{\omega}_{j}}}{\partial t}+\big( 16i\varepsilon\lambda_{j}^{5}+i\lambda_{j}^{2}\big)\Lambda {{\omega}_{j}} \right)=0,$$
which yields
$${{\omega}_{j}}={{{e}}^{\left(-i{{\lambda}_{j}}x-\left(16i\varepsilon \lambda_{j}^{5}+i\lambda_{j}^{2}\right)t\right)\Lambda}}{{\omega}_{j0}},\quad 1\le j\le N,$$
with ${\omega}_{j0}$ being independent of $x$ and $t$. In view of (24), we thus have
$${{\hat{\omega}}_{j}}=\omega_{j0}^{\dagger }{{{e}}^{\left(i\lambda_{j}^{*}x+\left(16i\varepsilon \lambda {{_{j}^{*}}^{5}}+i\lambda {{_{j}^{*}}^{2}}\right)t\right)\Lambda}},\quad 1\le j\le N.$$

In order to derive soliton solutions explicitly, we take $G=\mathbf{I}_{2}$ in (17), which indicates that no reflection exists in the scattering problem. Therefore, the solutions [24] for this special Riemann-Hilbert problem can be given by
\begin{subequations}
\begin{align}
&{{P}_{1}}(\lambda)=\mathbf{I}_{2}-\sum\limits_{k=1}^{N}{\sum\limits_{j=1}^{N}{\frac{{{\omega}_{k}}{{{\hat{\omega}}}_{j}}{{\big({{M}^{-1}}\big)_{kj}}}}{\lambda -{{{\hat{\lambda}}}_{j}}}}},\label{4.5.10a} \\
&
{{P}_{2}}(\lambda)=\mathbf{I}_{2}+\sum\limits_{k=1}^{N}{\sum\limits_{j=1}^{N}{\frac{{{\omega}_{k}}{{{\hat{\omega}}}_{j}}{{\big({{M}^{-1}}\big)_{kj}}}}{\lambda-{{\lambda }_{k}}}}},\label{4.5.10b}
\end{align}
\end{subequations}
with $M$ being defined by
$$
{{m}_{kj}}=\frac{{\hat{\omega}_{k}}{{{{\omega}}}_{j}}}{{{\lambda}_{j}}-{{{\hat{\lambda}}}_{k}}},\quad 1\le k,j\le N.
$$

In what follows, we intend to present reconstruction formula of the potential.
Because $P_{1}(x,\lambda)$ satisfies (3), we insert the expansion
$$
{{P}_{1}}(x,\lambda)={\mathbf{I}}_{2}+{\lambda}^{-1}{{P}_{1}^{(1)}}+{\lambda}^{-2}{{P}_{1}^{(2)}}+O\big({\lambda}^{-3}\big),\quad \lambda\to\infty.
$$
into (3) and generate
$$
Q=i\big[\Lambda,{P}_{1}^{(1)}\big]=\left( \begin{matrix}
   0 & 2i{{\big(P_{1}^{(1)}\big)_{12}}}  \\
   -2i{{\big(P_{1}^{(1)}\big)_{21}}} & 0  \\
\end{matrix} \right),
$$
from which we see
\begin{equation}
q=2i{{\big(P_{1}^{(1)}\big)_{12}}},
\end{equation}
where $\big(P_{1}^{(1)}\big)_{12}$ denotes the $(1,2)$-element of $P_{1}^{(1)}$.

Through supposing that
${{\omega}_{j0}}=(\alpha_{j},\beta_{j})^{\text{T}},\theta_{j}=-i{{\lambda}_{j}}x-(16i\varepsilon \lambda_{j}^{5}+i\lambda_{j}^{2})t$, therefore we acquire
the multi-soliton solutions of Eq. (1) as follows
\begin{equation}
q=-2i\sum\limits_{k=1}^{N}{\sum\limits_{j=1}^{N}{\alpha_{k}{{\beta_{j} }^{*}}{{{e}}^{{{\theta }_{k}}-\theta _{j}^{*}}}{{\big({{M}^{-1}}\big)_{kj}}}}},
\end{equation}
where
$$
{{m}_{kj}}=\frac{\alpha _{k}^{*}{{\alpha }_{j}}{{{e}}^{\theta _{k}^{*}+{{\theta }_{j}}}}+\beta _{k}^{*}{{\beta }_{j}}{{{e}}^{-\theta _{k}^{*}-{{\theta }_{j}}}}}{{{\lambda}_{j}}-\lambda_{k}^{*}},\quad 1\le k,j\le N.
$$

A selection of $N=1$ in (27) generates one-soliton solution as
\begin{equation*}
q=-\frac{2i\alpha _{1}{{\beta }_{1}^{*}}({{\lambda}_{1}}-\lambda_{1}^{*}){{{e}}^{{{\theta }_{1}}-\theta _{1}^{*}}}}{{{\left| {{\alpha }_{1}} \right|}^{2}}{{{e}}^{\theta _{1}^{*}+{{\theta }_{1}}}}+{{\left| {{\beta }_{1}} \right|}^{2}}{{{e}}^{-\theta _{1}^{*}-{{\theta }_{1}}}}},
\end{equation*}
which can be simplified into the form
\begin{equation}
q=2\alpha _{1}{{b}_{1}}{{{e}}^{-{{\xi }_{1}}}}{{{e}}^{\theta _{1}-{{\theta }_{1}^{*}}}}\text{sech}\left(\theta _{1}^{*}+{{\theta }_{1}}+{{\xi }_{1}}\right)
\end{equation}
due to the assumptions ${{\beta }_{1}}=1,{{\lambda}_{1}}={{a}_{1}}+i{{b}_{1}},{{\left| {{\alpha }_{1}} \right|}^{2}}={{{e}}^{2{{\xi }_{1}}}},$ and $\theta_{1}=-i{{\lambda}_{1}}x-(16i\varepsilon \lambda_{1}^{5}+i\lambda_{1}^{2})t$.

For $N=2$ in (27), two-soliton solution of Eq. (1) can be written as
\begin{equation}
q=-\frac{2i\big(\alpha _{1}{{\beta }_{1}^{*}}{{m}_{22}}{{{e}}^{{{\theta }_{1}}-\theta _{1}^{*}}}-\alpha _{1}{{\beta }_{2}^{*}}{{m}_{12}}{{{e}}^{{{\theta }_{1}}-\theta _{2}^{*}}}-\alpha _{2}{{\beta }_{1}^{*}}{{m}_{21}}{{{e}}^{{{\theta }_{2}}-\theta _{1}^{*}}}+\alpha _{2}{{\beta }_{2}^{*}}{{m}_{11}}{{{e}}^{{{\theta }_{2}}-\theta _{2}^{*}}}\big)}{{{m}_{11}}{{m}_{22}}-{{m}_{12}}{{m}_{21}}},
\end{equation}
where
$$\begin{aligned}
&{{m}_{11}}=\frac{{{\left| {{\alpha }_{1}} \right|}^{2}}{{{e}}^{\theta _{1}^{*}+{{\theta }_{1}}}}+{{\left| {{\beta }_{1}} \right|}^{2}}{{{e}}^{-\theta _{1}^{*}-{{\theta }_{1}}}}}{{{\lambda}_{1}}-\lambda_{1}^{*}},\quad
{{m}_{12}}=\frac{\alpha _{1}^{*}{{\alpha }_{2}}{{{e}}^{\theta _{1}^{*}+{{\theta }_{2}}}}+\beta _{1}^{*}{{\beta }_{2}}{{{e}}^{-\theta _{1}^{*}-{{\theta }_{2}}}}}{{{\lambda}_{2}}-\lambda _{1}^{*}}, \\
&{{m}_{21}}=\frac{\alpha _{2}^{*}{{\alpha }_{1}}{{{e}}^{\theta _{2}^{*}+{{\theta }_{1}}}}+\beta _{2}^{*}{{\beta }_{1}}{{{e}}^{-\theta _{2}^{*}-{{\theta }_{1}}}}}{{{\lambda}_{1}}-\lambda_{2}^{*}},\quad
{{m}_{22}}=\frac{{{\left| {{\alpha }_{2}} \right|}^{2}}{{{e}}^{\theta _{2}^{*}+{{\theta }_{2}}}}+{{\left| {{\beta }_{2}} \right|}^{2}}{{{e}}^{-\theta _{2}^{*}-{{\theta }_{2}}}}}{{{\lambda}_{2}}-\lambda_{2}^{*}}.
\end{aligned}$$
Under the assumptions ${{\beta }_{1}}={{\beta }_{2}}=1,{{\alpha }_{1}}={{\alpha }_{2}}$ and ${{\left| {{\alpha }_{1}}\right|}^{2}}={{{e}}^{2{{\xi }_{1}}}}$,
the solution (29) takes the form
\begin{equation*}
q=-\frac{2i\big(\alpha _{1}{{m}_{22}}{{{e}}^{{{\theta }_{1}}-\theta _{1}^{*}}}-\alpha _{1}{{m}_{12}}{{{e}}^{{{\theta }_{1}}-\theta _{2}^{*}}}-\alpha _{2}{{m}_{21}}{{{e}}^{{{\theta }_{2}}-\theta _{1}^{*}}}+\alpha _{2}{{m}_{11}}{{{e}}^{{{\theta }_{2}}-\theta _{2}^{*}}}\big)}{{{m}_{11}}{{m}_{22}}-{{m}_{12}}{{m}_{21}}},
\end{equation*}
where
\begin{align*}
&{{m}_{11}}=-\frac{i}{{{b}_{1}}}{{{e}}^{{{\xi }_{1}}}}\cosh\left(\theta _{1}^{*}+{{\theta }_{1}}+{{\xi }_{1}}\right),\quad {{m}_{12}}=\frac{2{{{e}}^{{{\xi }_{1}}}}}{({{a}_{2}}-{{a}_{1}})+i({{b}_{1}}+{{b}_{2}})}\cosh \left(\theta _{1}^{*}+{{\theta }_{2}}+{{\xi }_{1}}\right),\\
&{{m}_{22}}=-\frac{i}{{{b}_{2}}}{{{e}}^{{{\xi }_{1}}}}\cosh\left(\theta _{2}^{*}+{{\theta }_{2}}+{{\xi }_{1}}\right),\quad {{m}_{21}}=\frac{2{{{e}}^{{{\xi }_{1}}}}}{({{a}_{1}}-{{a}_{2}})+i({{b}_{1}}+{{b}_{2}})}\cosh \left(\theta _{2}^{*}+{{\theta }_{1}}+{{\xi }_{1}}\right).
\end{align*}

\section{Conclusion}
The aim of the current research was to work out multi-soliton solutions of the quintic nonlinear Schr\"odinger equation. For this purpose, we first carried out the spectral analysis and formulated the related matrix Riemann-Hilbert problem on the real line. Second, based on the resulting Riemann-Hilbert problem which was treated by considering that no reflection exists in the scattering problem, the general multi-soliton solutions for the quintic nonlinear Schr\"odinger equation were generated in explicit form. Particularly, the one- and two-soliton solutions were given.


\newpage

\end{CJK*}  
\end{document}